\documentclass[twocolumn]{aastex61}

\usepackage{ulem}

\newcommand{\iso}[2]{\hbox{${}^{#1}{\rm #2}$}}

\received{November 6, 2017}
\revised{February 15, 2018}
\accepted{February 27, 2018}

\shorttitle{A formation timescale of the Galactic halo}
\shortauthors{Carlos et al.}

\begin{document}

\title{A formation timescale of the Galactic halo from Mg isotopes in dwarf stars\footnote{The data presented herein were obtained at the W.M. Keck Observatory, which is operated as a scientific partnership among the California Institute of Technology, the University of California and the National Aeronautics and Space Administration. The Observatory was made possible by the generous financial support of the W.M. Keck Foundation.}}

\correspondingauthor{Mar\'ilia Carlos}
\email{marilia.carlos@usp.br}

\author[0000-0003-1757-6666]{Mar\'ilia Carlos}
\affil{Universidade de S\~ao Paulo, IAG, Departamento de Astronomia \\
Rua do Mat\~ao 1226, Cidade Universit\'aria, 05508-900, S\~ao Paulo, SP, Brazil}
\affil{Monash Centre for Astrophysics, School of Physics and Astronomy, Monash University\\
Victoria 3800, Australia}

\author{Amanda I. Karakas}
\affiliation{Monash Centre for Astrophysics, School of Physics and Astronomy, Monash University\\
Victoria 3800, Australia}

\author{Judith G. Cohen}
\affiliation{California Institute of Technology \\
1200 E. California Boulevard, MC 249-17, Pasadena, CA 91195, USA}

\author{Chiaki Kobayashi}
\affiliation{Centre for Astrophysics Research, School of Physics, Astronomy and Mathematics, University of Hertfordshire \\
College Lane, Hatfield AL10 9AB, UK}

\author{Jorge Mel\'endez}
\affil{Universidade de S\~ao Paulo, IAG, Departamento de Astronomia \\
Rua do Mat\~ao 1226, Cidade Universit\'aria, 05508-900, S\~ao Paulo, SP, Brazil}



\begin{abstract}

We determine magnesium isotopic abundances of metal poor dwarf stars from the galactic halo, to shed light on the onset of asymptotic giant branch (AGB) star nucleossynthesis in the galactic halo and constrain the timescale of its formation.
We observed a sample of eight new  halo K dwarfs in a metallicity range of $-1.9 < \mathrm{[Fe/H]} < -0.9$ and 4200 $< T_{\mathrm{eff}}\mathrm{(K)} <$ 4950, using the HIRES spectrograph  at the Keck Observatory ($R\approx10^5$ and $200 \leq \mathrm{S/N} \leq 300$). We obtain magnesium isotopic abundances by spectral synthesis on three MgH features and compare our results with galactic chemical evolution models.
With the current sample, we almost double the number of metal poor stars with  Mg isotopes determined from the literature. The new data allow  us to determine the metallicity when the \iso{26}Mg abundances start to became important, $\mathrm{[Fe/H]} \sim -1.4 \pm 0.1$. The data with $\mathrm{[Fe/H]} > -1.4$ are somewhat higher (1-3 $\sigma$) than previous chemical evolution model predictions, indicating perhaps higher yields of the neutron-rich isotopes. Our results using only AGB star enrichment suggest a timescale for formation for the galactic halo of about 0.3 Gyr, but considering also supernova enrichment, the upper limit for the timescale formation is about 1.5 Gyr.

\end{abstract}




\section{Introduction}
\label{intro}

The study of galaxy formation and evolution is a vigorous field of astronomy, with many studies in the literature debating how our Galaxy evolved dynamically and chemically. Regarding the studies of the evolution of the Galactic halo, an important uncertainty is its formation timescale. The classic work of \cite{egen62} discusses a monolithic scenario in which a fast dissipative collapse occurs on a timescale of 200 million years. Later, \cite{searly78} suggested a central collapse, but implied that the outer halo was formed by the merging of larger fragments, resulting in a formation timescale $>1$ Gyr. The latter approach is similar to current cosmological $\Lambda$CDM models in which larger galaxies, such as the Milk Way, were formed hierarchically (e.g., \citealt{navarro97} and \citealt{zolotov/09}). Hierarchical chemico-dynamical models show that 80\% of the galactic halo has [O/Fe]$\gtrsim0.5$ \cite[Figure 9]{kobayashi/nakasato/11} and \cite{tissera/12} indicate the presence of accreted stars with high $\alpha-$enhancement in the outer region of the galactic halo. Those hierarchical models  are in agreement with observations that indicate that there are at least two different stellar populations in the halo (e.g., \citealt{carollo07,nissen10}).

Assuming different gas infall episodes that contribute to the formation of the galactic components, chemical evolution models for the Galaxy, such as those by \cite{chiappini97} and \cite{micali13}, put a constraint on the timescales of star formation and chemical enrichment of the components. These types of models find values for the formation timescale of the halo that vary from 0.2 to 2 Gyr, where the gas accretion rate depends on the formation timescale of the halo, and thin and thick disk components, which will set the metallicity distribution function of the galaxy. Thus, knowing the timescale for formation of the various components of the Galaxy can constrain chemical evolution models.
 
Elemental and isotopic abundances from different nucleosynthetic sites are an extremely useful tool to solve this problem. Since the different isotopes could be formed in stars of different masses, which die at different ages, they could function as  ``clocks'' to trace the timescales of halo formation.

Magnesium in particular is a good clock because its different isotopes are produced in different sites (i.e., different stars); therefore, they trace stellar (and Galactic) evolution over short and long timescales. The element magnesium  has three stable isotopes: \iso{24}Mg, \iso{25}Mg and \iso{26}Mg. The magnesium isotopes \iso{24,25,26}Mg are produced inside massive stars, while the isotopes \iso{25,26}Mg  are also produced  in stars with intermediate mass (we discuss the details of Mg production in Sect. \ref{mgh_prod}). Since the \iso{25,26}Mg isotopes can be produced in  asymptotic giant branch (AGB) stars \citep{karakas03,fishlock14}, measuring the Mg isotopic ratios can inform us when the heavier Mg isotopes from AGB stars begin to contribute toward galactic chemical enrichment, meaning that the isotopic ratios \iso{25,26}Mg/\iso{24}Mg increase with the onset of AGB stars.

Some chemical evolution models include the chemical abundances of magnesium and its stable isotopes (e. g. \citeauthor{alibes2001} \citeyear{alibes2001}; \citeauthor{fenner03} \citeyear{fenner03} and \citeauthor{kobayashi11} \citeyear{kobayashi11}). Despite the existence of these models, there are few observations and analysis of these isotopes that could indicate which model is more appropriate. It is also important to stress that isotopic abundances offer more observational links than elemental abundances, because several specific nucleosynthetic processes produce isotopes, while the elemental abundance is the sum of all the isotopes that compose an element.

There are several important contributions in the literature regarding the determination of Mg isotopic abundances, such as \cite{barbuy85}, \cite{barbuy87a}, \cite{barbuy87b}, \cite{gay00}, \cite{yong/grundahl03}, \cite{yong2003}, \cite{yong/lambert04}, \cite{melendez07}, \cite{melendez09} and \cite{thygesen16},  but due to the difficulty in measuring the MgH lines we have little data, especially at low metallicities,  to assess the evolution of the Mg isotopic ratios. Note that the MgH features are visible only in cool and not too evolved stars \citep{spinrad65}. Regarding the halo population, a limited number of halo stars have been analyzed, with important contributions by \cite{yong2003} and \cite{melendez07}. Only seven single metal poor stars ($-2.60<$ [Fe/H] $<-1.35$) from the Galactic halo have Mg isotopic measurements in the literature.   Therefore, in this paper, we extended the  metallicity range  with our new sample and make it possible to assess when the \iso{25,26}Mg abundances start to became important with respect to \iso{24}Mg abundances to constrain the onset of AGB stars in the galactic halo, adding more insights to the galactic chemical evolution process.

The paper is organized as follows: in Sect. \ref{mgh_prod} we discuss how the three magnesium isotopes are formed; in Sect. \ref{obs}, we show the sample and the stellar parameters; in Sect. \ref{analysis}, we describe the analysis, Sect. \ref{discussion} shows the results and discussion and the conclusions are presented in Sect. \ref{conclusion}.


\section{Production of Magnesium isotopes in different stellar sites}
\label{mgh_prod}

The main magnesium isotope \iso{24}Mg is produced inside massive stars during core carbon and neon burning before the supernova explosion. During core carbon burning, one of the most important reactions is $^{12}\rm{C}(^{12}\rm{C,p})^{23}\rm{Na}$, where the product \iso{23}Na is destroyed through the reaction $^{23}\rm{Na}(\rm{p},\gamma)^{24}\rm{Mg}$ which is responsible for the creation of \iso{24}Mg. According to \cite{arnett/85}, \iso{24}Mg  is the third most important product of  core carbon burning in massive stars. During core neon burning, $^{20}\rm{Ne}(\gamma,\alpha)^{16}\rm{O}$ is the main reaction \citep{thielemann/85} which generates $\alpha-$particles. These $\alpha-$particles, along with the remaining \iso{20}Ne, form \iso{24}Mg through the reaction $^{20}\rm{Ne}(\alpha,\gamma)^{24}\rm{Mg}$.

The isotopes \iso{25,26}Mg are also produced in smaller amounts in massive stars in their outer carbon layers during helium burning \citep{ww95} through the reactions $^{22}\rm{Ne}(\alpha,\rm{n})^{25}\rm{Mg}$, $^{22}\rm{Ne}(\alpha,\gamma)^{26}\rm{Mg}$  and $^{25}\rm{Mg}(\rm{n},\gamma)^{26}\rm{Mg}$. For details on the amount of \iso{24,25,26}Mg produced in massive stars, see \cite{heger/10}.

The magnesium isotopes are additionally produced in AGB stars. They can be formed in three possible regions: the hydrogen burning shell, the helium burning shell, and at the base of the convective envelope in intermediate-mass stars during hot bottom burning (for a review of AGB evolution, see \citealt{karakas14}).

The most important production site of \iso{25,26}Mg is the helium burning shell in AGB stars \citep{karakas03}. During a thermal pulse, \iso{22}Ne is created via successive $\alpha$-captures onto \iso{14}N. When the temperature of this region increases above $\approx 300 \times 10^6$ K, the stars experience an increase in \iso{25,26}Mg through the reactions $^{22}\rm{Ne}(\alpha,\rm{n})^{25}\rm{Mg}$ and $^{22}\rm{Ne}(\alpha, \gamma)^{26}\rm{Mg}$. Additionally, \iso{26}Mg may also be produced by neutron capture via \iso{25}Mg$(n,\gamma)$\iso{26}Mg.

During hot bottom burning in massive AGB stars, hydrogen burning occurs via the CNO cycle, Ne -- Na and Mg -- Al chains when the temperature is higher than $50\times10^6$ K. The lower densities at the base of the envelope mean that hydrogen burning reactions need to occur at higher temperatures than described in, e.g., \cite{arnould/99}. Thus, this site becomes important for the production and depletion of the magnesium isotopes (see \citealt{karakas03} and \citealt{ventura11}).

Magnesium isotopes are created via the Mg -- Al chain (for details, see \citealt{arnould/99} or \citealt{karakas03}), but in the same chain they can also be destroyed. The isotope \iso{25}Mg is destroyed through the reaction $^{25}\rm{Mg}(\rm{p},\gamma)^{26}\rm{Al}$. The isotope \iso{26}Mg experiences a little decrement through the reaction $^{26}\rm{Mg}(\rm{p},\gamma)^{27}\rm{Al}$ until temperatures of  $\approx 60 \times 10^6$ K, but also experiences an abundance enhancement due to the decay of \iso{26}Al in the hydrogen shell ashes. The abundance of \iso{24}Mg remains almost stable at temperatures below about $70\times10^6$ K in the hydrogen burning shell, but in the most massive AGB stars the temperature at the base of the envelope may exceed $90\times10^6$ K, hot enough for the destruction of \iso{24}Mg by proton capture.

Altogether, AGB stars are responsible for a considerable amount of \iso{25,26}Mg isotopes produced in the Galaxy. Since the lifetime of a star depends on its mass and metallicity, the study of Mg abundances in Galactic halo main-sequence stars, which do not have their chemical composition affected by stellar evolution, can determine the onset of the effects of AGB evolution in the Galactic halo, and this can provide us insights on the timescale for formation of the Galactic halo.


\section{Spectra and stellar parameters}
\label{obs}

The sample consists of  eight  K dwarf stars from the galactic halo. These objects were chosen from the updated catalog of \cite{ramirez05}, where we considered the temperature interval (4000-5000 K) as well as the metallicity range of $-2< \mathrm{[Fe/H]}  <-0.8$. We discard binary stars to avoid contamination from the companion.

In order to get precise measurements for the three Mg isotopes, we need high resolution and good signal-to-noise spectra \textbf{($>150$)}. These conditions were achieved thanks to the HIRES spectrograph \citep{hires94} at the Keck Observatory ($R\approx10^5$ and $200 \leq \mathrm{S/N} \leq 300$) in 2007 September. The spectral orders were extracted with MAKEE\footnote{MAKEE was developed by T. A. Barlow specifically for reduction of Keck HIRES data. It is freely available at \url{http://www.astro.caltech.edu/\~tb/makee/}.}. For Doppler correction, combining spectra, and continuum normalization, we used IRAF\footnote{\url{http://iraf.noao.edu/}.}.

After the data reduction, we found that there were two double-lined stars in our sample, BD -004470 and G 3-13, which were discarded from the analysis. For the remaining stars, the stellar effective temperatures were derived according to the photometric calibration from \cite{casagrande10}, using the values of {\it B}, {\it V}, {\it J}, {\it H}, and {\it Ks} magnitudes compiled by \cite{pastel16}. 

The [Fe/H] and microturbulence values were determined by measuring Fe {\sc I} and Fe {\sc II} lines with the aid of IRAF and using the  2014 July version of the 1D LTE code MOOG \citep{sneden73}. The Fe {\sc I} and Fe {\sc II} line list specifically for metal poor K dwarfs was taken from the work of \cite{chen06}. The T$_{\mathrm{eff}}$ values derived here are compatible with previous works in the literature. We adopted literature values of surface gravity (log g) (\citeauthor{ramirez05} \citeyear{ramirez05}; \citeauthor{yong03pasp} \citeyear{yong03pasp}). The stellar parameters are presented in Table \ref{tab1}. 

As a result of the abundance analysis, we found a star with chemical anomalies, named LHS 173. The analysis of that star will be presented elsewhere.

\section{Analysis}
\label{analysis}

Since the isotopes $^{25,26}$Mg have a weak contribution in the wings of a stronger $^{24}$MgH line, creating a red asymmetry in the MgH feature, we have to employ spectral synthesis to derive the Mg isotopic abundances.

\begin{figure}
   \centering
   \includegraphics[width=\hsize]{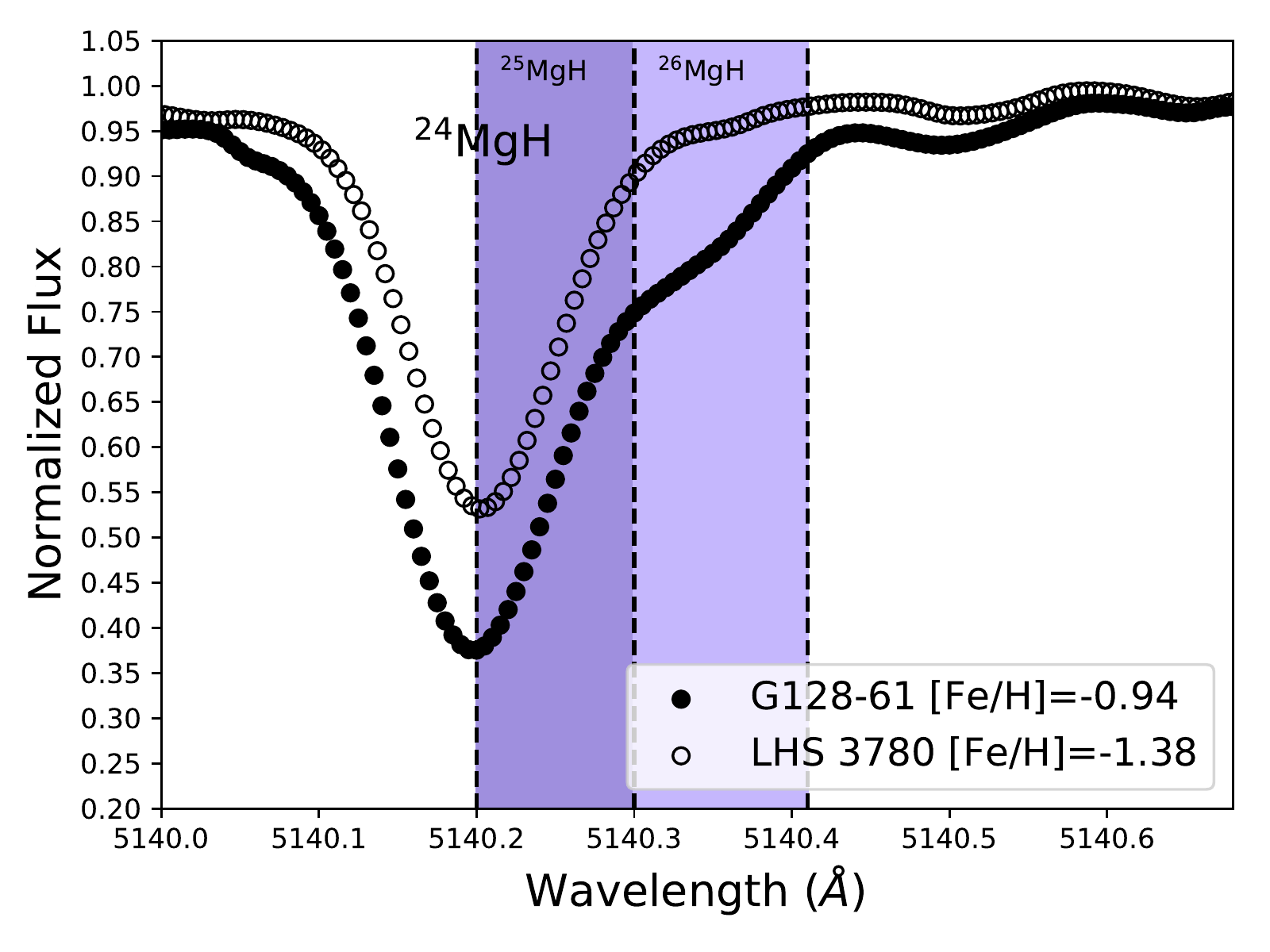}
      \caption{Spectra from stars with different [Fe/H] in the region  5140.2 \AA \, of the MgH molecular feature.
              }
         \label{fig1}
   \end{figure}

As we can see in Fig. \ref{fig1}, for stars with similar temperatures but different [Fe/H], we have different amounts of MgH. The more metal poor the star is, the less MgH   will be present. Furthermore, we can see that the red asymmetry is stronger in the more metal-rich star, suggesting a higher fraction of $^{25,26}$Mg. 

We determined the macroturbulence velocity broadening by analyzing the line profiles of the Fe {\sc I} 6056.0 \AA, 6078.5 \AA, 6096.7 \AA \, and 6151.6 \AA \, lines, setting the rotational velocity broadening as zero. We also included the instrumental broadening in the calculations.

As recommended by \cite{mcwilliam88} and \cite{gay00}, and also used in the works of \citeauthor{barbuy85} (\citeyear{barbuy85}, \citeyear{barbuy87a}), \cite{yong2003}, \cite{melendez07} and \cite{melendez09}, we adopted three wavelength regions to determine the Mg isotopic abundances ratios, namely 5134.6 \AA, 5138.7 \AA \, and 5140.2 \AA.

The isotopic abundances are estimated as described in \cite{melendez07}, also using the code MOOG and the Kurucz grid of ATLAS9 model atmospheres \citep{castelli/04}. The line list adopted in this region is the same used in the work of \cite{melendez07}, which includes both molecular and atomic lines. Although  the abundance analysis was derived with 1D models, \cite{thygesen16,thygesen17} showed that using 3D models in the analysis of Mg isotopes does not have a significant impact, especially for \iso{26}Mg/Mg.

\begin{figure*}
\begin{tabular}{c c}

\includegraphics[width=0.5\textwidth]{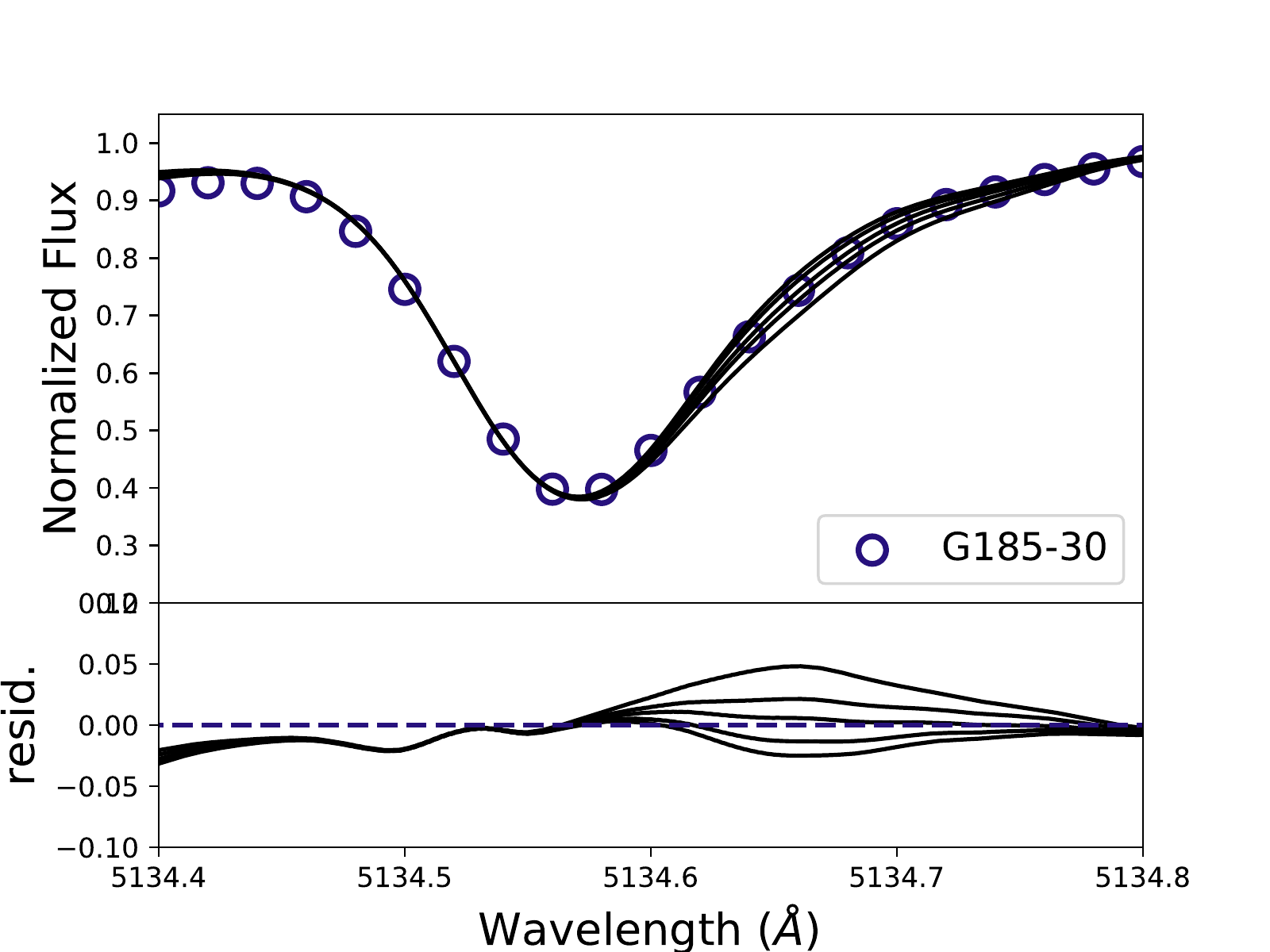} & \includegraphics[width=0.5\textwidth]{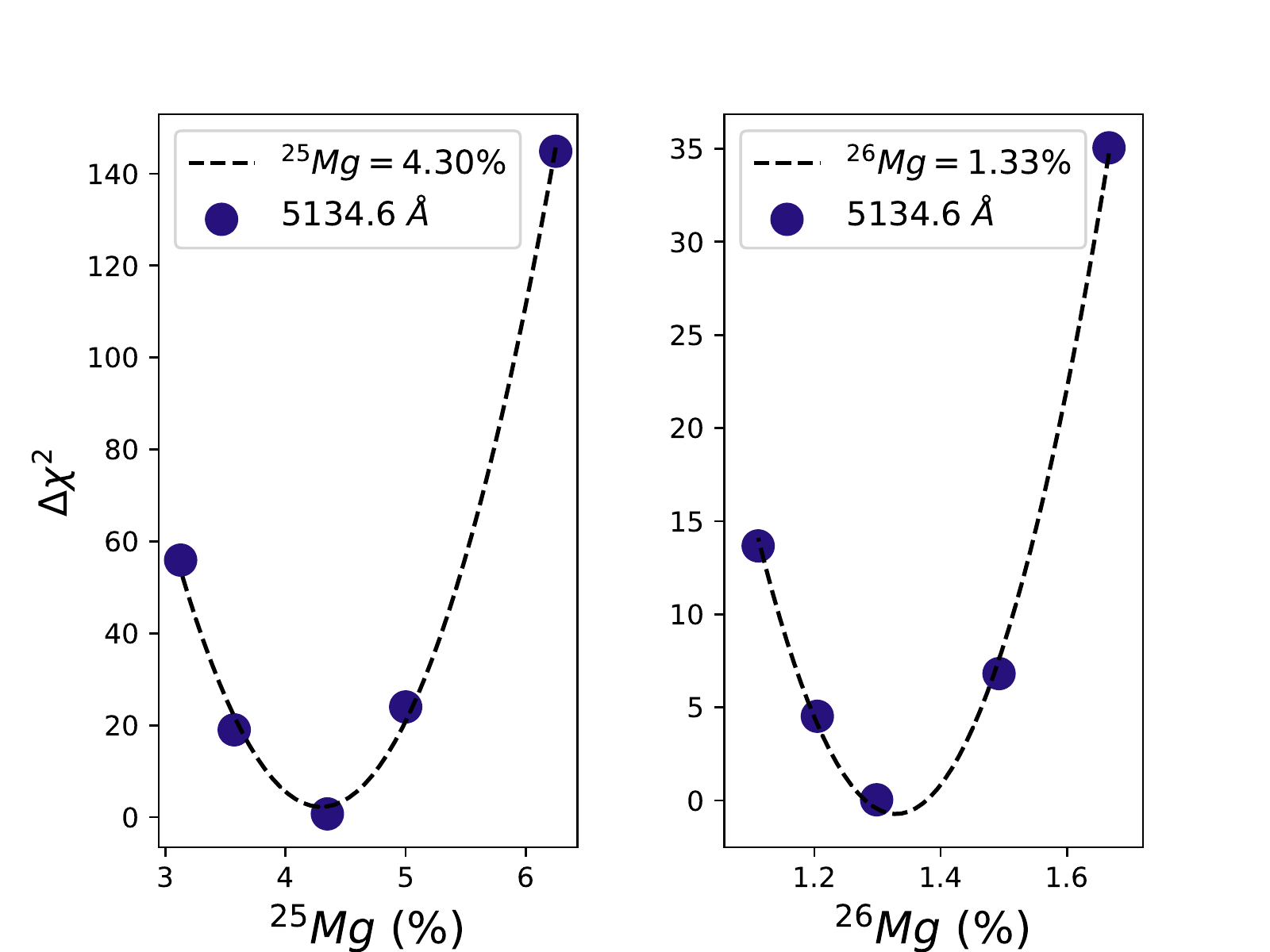} \\
\end{tabular}

\caption{Left panel shows the MgH 5134.6 \AA \, region with the observed spectrum (blue open circles) and the respective spectral synthesis for five different values represented by black solid lines. The right panel shows the $\chi^2$ analysis with the best value of $^{25}$Mg and $^{26}$Mg. Both panels are for the star G 185-30.}
\label{fig2}

\end{figure*}

The isotopic abundances are measured by performing a $\chi^2$ fit, where $\chi^2=\Sigma(O_i-S_i)/\sigma^2$, with $O_i$ and $S_i$ being the observed and synthetic spectra and $\sigma=(S/N)^{-1}$. A comparison of spectral synthesis and observed spectra is shown in the left panel of Fig. \ref{fig2}. The right panel of the Fig. \ref{fig2} displays the variations of the $\chi^2$ fits.

The final isotopic values, as well as the parameters adopted in the spectral synthesis, are presented in Table \ref{tab1}. The \iso{25,26}Mg errors are the standard deviation between the isotopic ratios of the three regions adopted in this work. The line-to-line scatter in the isotopic percentages is only about 1\%, showing the consistency among the different MgH features and the high precision achieved in this work.

\begin{table*}
\caption{Stellar parameters and magnesium isotopic ratios.}             
\label{tab1}      
\centering                          
\begin{tabular}{l c c c c r r}        

\hline                
\noalign{\smallskip}
Object & T$_{\mathrm{eff.}}$ (K) & [Fe/H] & log g (dex) & v$_{\mathrm{mic.}}$ (km s$^{-1}$)& $^{25}$Mg (\%) & $^{26}$Mg (\%)\\    
\noalign{\smallskip}
\hline\hline                         
\noalign{\smallskip}
   G 185-30 & 4524 & -1.85$\pm$0.01 & 4.5$^{a}$ & 0.00 & \textcolor{black!50}{4.0$\pm$0.3}  & 1.6$\pm$0.4\\ 
   G 128-61 & 4664 & -0.94$\pm$0.02 & 5.0$^{b}$ & 0.00 & \textcolor{black!50}{8.0$\pm$1.0} & 4.8$\pm$1.6\\
   G 78-26 & 4288 & -1.20$\pm$0.02  & 4.7$^{b}$ & 0.24 & \textcolor{black!50}{5.3$\pm$0.2} & 3.4$\pm$0.6\\
   G 189-45 & 4937 & -1.33$\pm$0.01 & 4.3$^{b}$ & 0.00 & \textcolor{black!50}{4.6$\pm$1.1} & 2.2$\pm$0.9\\
   LHS 3780 & 4880 & -1.38$\pm$0.01 & 4.5$^{b}$ & 0.00 & \textcolor{black!50}{4.5$\pm$0.1} & 0.0$\pm$1.0\\ 
   Sun & 5777 & 0.00 & 4.44 & 1.00 & \textcolor{black!50}{10.00$^{c}$} & 11.01$^{c}$\\
\hline                                   
\end{tabular}
\begin{tabular}{l}
\textbf{Notes.} $^{(\ast)}$Magnesium isotopic ratios are given with respect to $^{24}$Mg + $^{25}$Mg + $^{26}$Mg. \\
\textbf{References.} $^{(a)}$\cite{ramirez05}. $^{(b)}$\cite{yong03pasp}. $^{(c)}$\cite{asplund09}.
\end{tabular}

\end{table*}

\section{Discussion}
\label{discussion}

\begin{figure*}
\begin{tabular}{c c}

\includegraphics[width=0.5\textwidth]{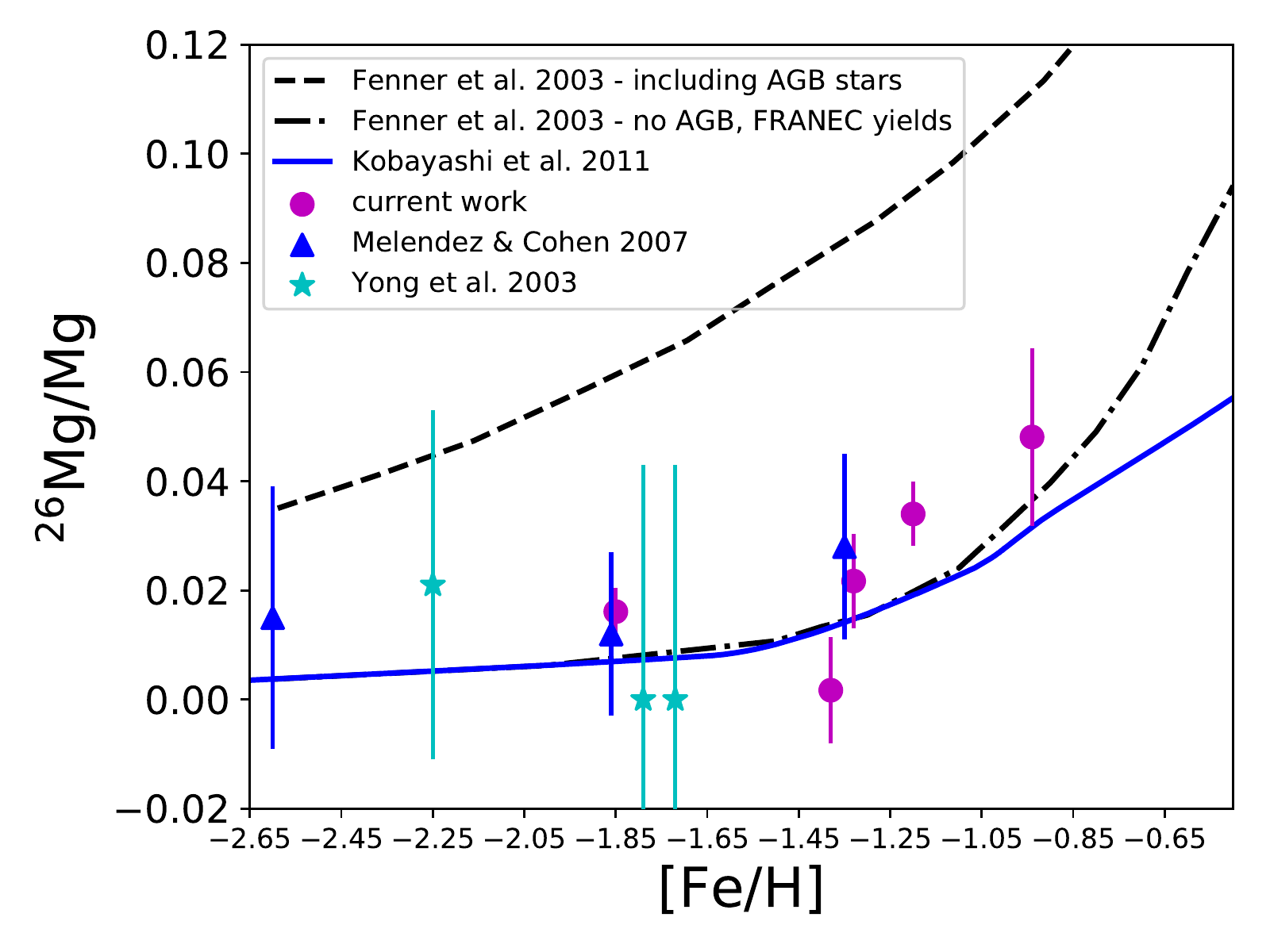} & \includegraphics[width=0.5\textwidth]{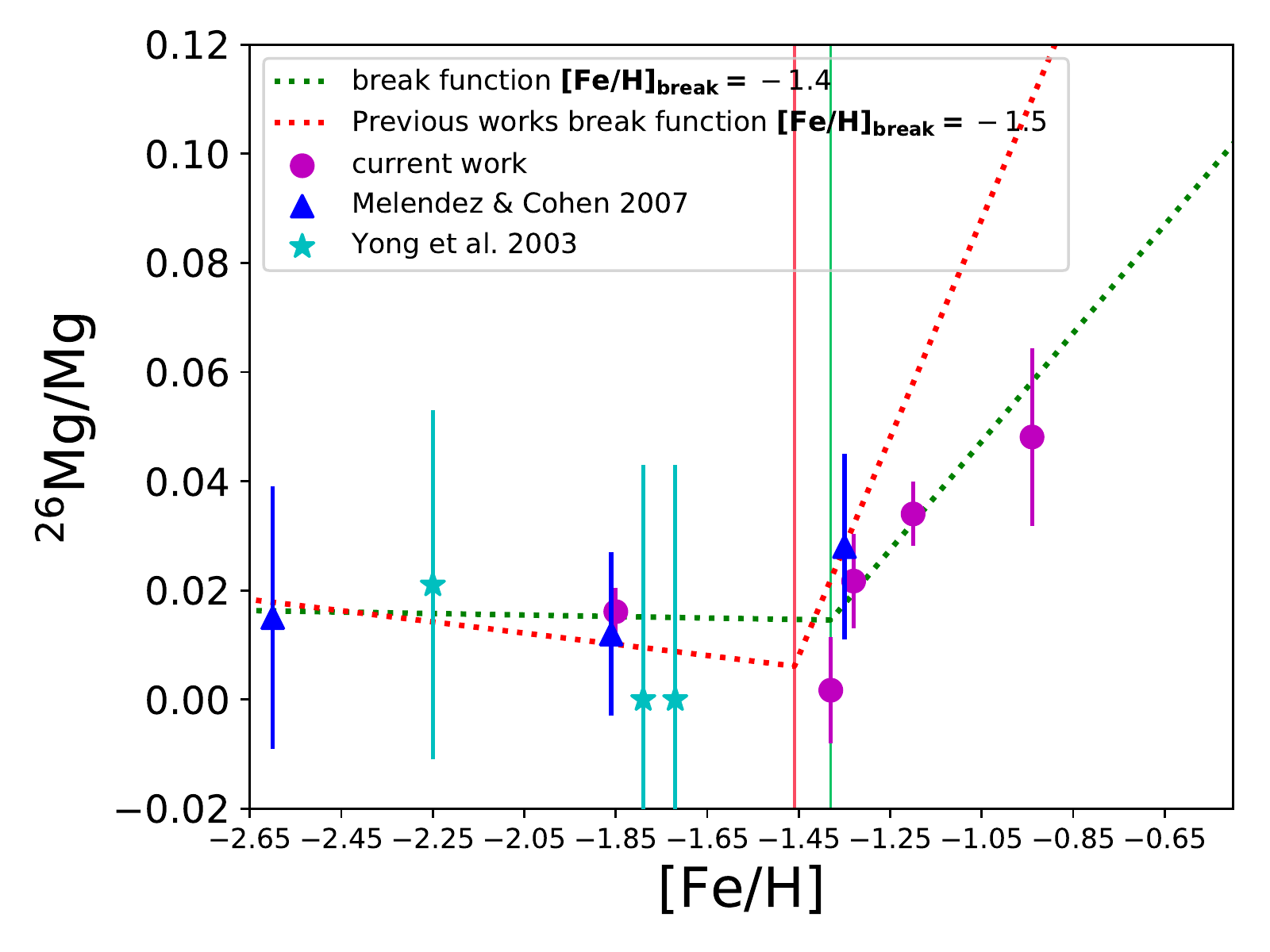} \\
\end{tabular}
      \caption{In both panels, purple circles represent the data determined in this work, blue triangles show the data from \protect\cite{melendez07} and the cyan stars exhibit results from \protect\cite{yong2003}. In the left panel, the black dotted-dashed line shows a model from \protect\cite{fenner03} with no AGB contribution, the black dashed line shows a model from the same work with AGB contribution and the model for the solar neighborhood from \protect\cite{kobayashi11} is represented by the blue solid line.  The green dotted line in the right panel shows the break function considering all the observed data and the red dotted line shows the break function considering only the data from \protect\cite{yong2003} and \protect\cite{melendez07}, the red and green vertical lines  indicate [Fe/H]$=-1.5$ and [Fe/H]$=-1.4$ respectively.}
         \label{fig3}
   \end{figure*}

The results including the current analysis plus data from the literature are shown in Fig. \ref{fig3}, note that the star LHS 3780 from the sample of \cite{yong2003} is not shown here since we present new isotopic abundances for this star in this work. The \iso{25}Mg/Mg ratios should not be used to compare with Galactic chemical evolution models due to both observational uncertainties arising from the smaller isotopic shift in comparison with \iso{24}Mg and to modeling uncertainties (i.e., the effects of 3D hydrodynamical model atmospheres shown in \citealt{thygesen16,thygesen17}). However, the  \iso{26}Mg/Mg ratio is robust, as the \iso{26}Mg determination is almost immune to the effects of 3D hydrodynamical model atmospheres, as discussed in \cite{thygesen16,thygesen17}.

We see that our data are consistent with previous measurements in the literature and with the model of \cite{fenner03}, which does not include AGB stars, and \cite{kobayashi11}, which includes the contribution from AGB stars,  for halo dwarfs with [Fe/H] $< -1.4$. However,  the model of \cite{fenner03} including the AGB contribution, does not match with the observational data.

It is possible to see as well that for stars with [Fe/H] $> -1.4$, the data differ somewhat (1-3 $\sigma$) from either of the models (\citealt{fenner03} without AGB stars, and \citealt{kobayashi11} including AGB stars). This suggests higher yields of the neutron-rich isotopes, in contrast to current yield predictions, or perhaps a different value for the timescale formation of the halo (see below). 

We adjusted a break function (Fig. \ref{fig3}, green dotted line in the right panel), including the data from the current work and values from the literature, in order to estimate a reliable metallicity at which low-metallicity AGB stars begin to  contribute to galactic chemical enrichment. The metallicity achieved with this work ([Fe/H]$=-1.4\pm0.1$) is slightly higher than the one determined by the study of \cite{melendez07}. According to our results, AGB stars begin to contribute to the Mg isotopes at a metallicity of [Fe/H] $> -1.4$. In order to compare our contribution with a new sample of stars relative to the previous data in the literature, we also adjusted a break function considering only the data from literature (Fig. \ref{fig3}, red dotted line in the right panel). Thus, we conclude that our new data is essential to better establish the break point when \iso{26}Mg/Mg starts to rise.

\begin{figure}
   \centering
   \includegraphics[width=\hsize]{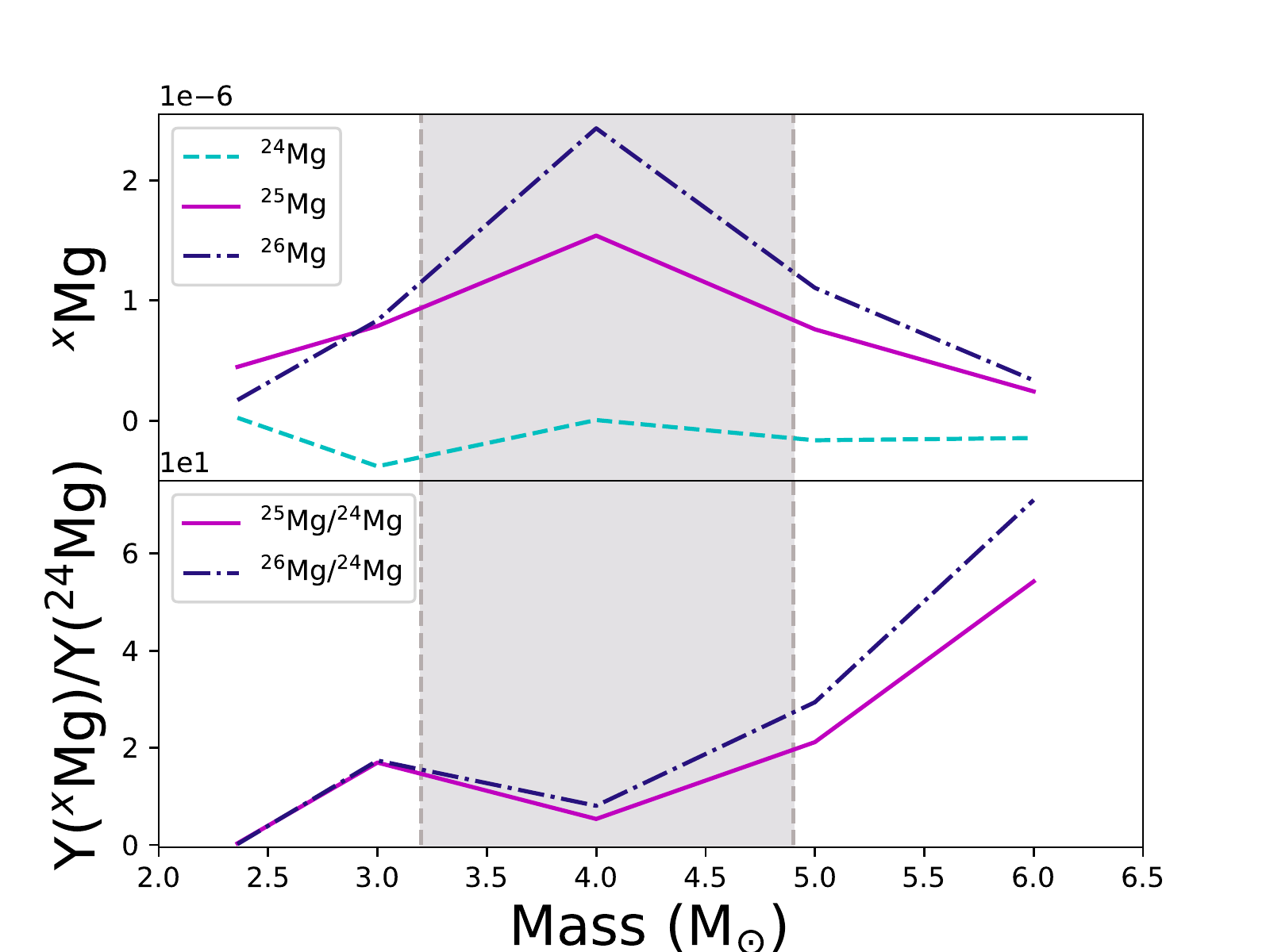}
      \caption{Top: AGB stellar yields from \protect\cite{shingles15} weighted by the initial mass function of \protect\cite{kroupa93}. Bottom:  mass isotopic ratios for the same data. The gray area indicates the region where we have the majority contribution of \iso{25,26}Mg isotopes. Data are from \protect\cite{shingles15}.
              }
         \label{fig4}
   \end{figure}

The study of \cite{shingles15} suggests that for [Fe/H] = -1.4 the majority of the contribution comes from AGB stars with $\gtrsim 4\pm1$ solar masses (Fig. \ref{fig4}). For stars in this mass interval, the lifetime is between approximately $\lesssim 150-300$ million of years. Thus, if we were just to consider AGB stars, we would suggest a short timescale for the formation of the galactic halo.

\begin{figure}
   \centering
   \includegraphics[width=\hsize]{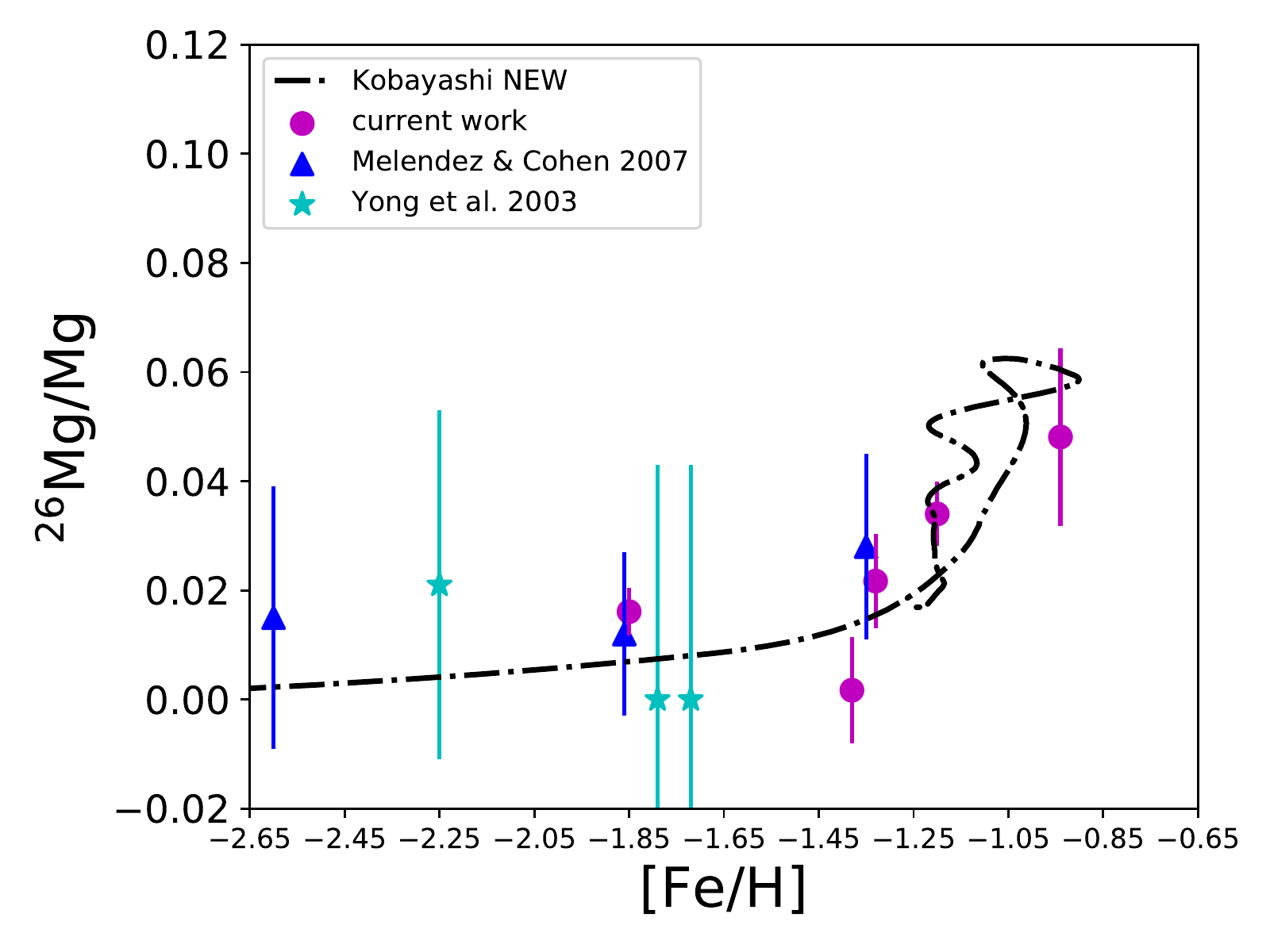}
      \caption{Our new model based upon \protect\cite{kobayashi11}, but now considering strong outflow. The observed data (see Fig. \protect\ref{fig3}) are also plotted for comparison.
              }
         \label{fig5}
   \end{figure}

However, if we simply set a short duration of star formation for the halo, it is not possible to reproduce our observed \iso{26}Mg/Mg ratios at [Fe/H] $> -1.4$. This is because core-collapse supernovae produce \iso{24}Mg at the same time when AGB stars produce \iso{26}Mg. It is necessary to suppress the contribution from core-collapse supernovae and to make the AGB contribution dominant for the chemical enrichment in the halo. One way to model this is to introduce a strong outflow. In Fig. \ref{fig5}, a new galactic evolution model (dotted-dashed line) considering this strong gas outflow for the halo is presented and compared to our data. In this new model, the chemical evolution in a system (i.e., Galactic halo) is numerically computed with the basic equations in \citealt{kobayashi00} (Equations 5 and 6) and the outflow term  in \citealt{kobayashi06} (page 1165). The SFR is proportional to the gas fraction; $\phi=(1/\tau_{\rm s})f_{\rm g}$. The driving source of the outflow is the feedback from supernovae, and hence the outflow rate is also proportional to the gas fraction; $R_{\rm out}=(1/\tau_{\rm o})f_{\rm g}$. The initial gas fraction is set to be $f_{\rm g}(0)=1$ with no metallicity. The new stars are formed from the mix of the remaining primordial gas plus any gas ejection of previous generations of stars (i.e., mass loss and supernovae). The outflow also removes some metals with the composition of the average metallicity of the system at the time, $R_{\rm out}Z(t)$. The outflow gas could later fall onto the disk, but this process is not included in the model. Since this is not a dynamical model, the timescales are determined to reproduce observations, namely, the observed metallicity distribution function (\citealt{chiba98}, see also \citealt{kobayashi11} for a more detailed discussion).  It is possible to have inflow as well, but the timescale should be short. Otherwise, it is not possible to reproduce the low metallicity of the Galactic halo stars.

In the best-fit model, the star formation and outflow timescales are $\tau_{\mathrm{s}}=5$ and  $\tau_{\mathrm{o}}=0.2$ Gyr, respectively, with no inflow. The Kroupa IMF is adopted. Note that super-AGB yields are also included in this model, but the contribution is negligible (C. Kobayashi et al. 2018, in preparation). Other parameter sets are also possible such as $\tau_{\rm s}=10$ and  $\tau_{\rm o}=0.4$ Gyr, but the outflow timescale should be no longer than $\tau_{\rm o}=0.4$ Gyr to explain our data at [Fe/H]$=-0.94$. Half of the halo stars are likely to be formed within 0.7 Gyr for $\tau_{\rm o} = 0.2$ Gyr or should be formed within 1.5 Gyr for $\tau_{\rm o} = 0.4$ Gyr. With these short outflow timescale, the outflow gas contains the metals mostly produced by supernovae, while the new stars contain the metals  mostly ejected from AGB stars. In the original halo model in \cite{kobayashi11} with $\tau_{\rm s}=15$ and  $\tau_{\rm o}=1$ Gyr, only 15\% of halo stars are formed within $1.5$ Gyr, and this model does not show the rapid increase of \iso{26}Mg/Mg ratios, very similar to the solar neighborhood model in Fig. \ref{fig3} (solid line). From the various constraints, we suggest that the timescale for the formation of the halo should be below 1.5 Gyr.

Although we used a one-zone chemical evolution model for comparison with the observed data, we can stress that our short timescales of inflow and outflow suggest that the potential of the star forming region is likely  shallow, and the progenitor system could be satellite galaxies, which is consistent with hydrodynamical simulations such as in \cite{monachesi/17}.

\section{Conclusions}
\label{conclusion}

Due to high resolution and excellent signal-to-noise spectra obtained with HIRES at the 10 m Keck I telescope, we were able to determine the magnesium isotopic abundances with very high precision for five stars, thus almost doubling the data from the literature of single metal poor halo dwarfs.

Here we stress that our conclusions are made with the addition of only two more stars at the high metallicity end, adding more is difficult due to observational limitations. It is with  this additional data that it is now possible to better estimate when the contribution from AGB stars became important for the galactic halo.  From this work, we can confirm that \iso{26}Mg abundances start to rise for stars with [Fe/H] $> -1.4$.

We conclude that for $\mathrm{[Fe/H]} > -1.4$ the Mg isotope ratios somewhat disagree with previous chemical evolution model predictions, indicating higher yields of the neutron-rich isotopes, in contrast to current yield predictions.  However, the disagreement between the data and the models can also be explained with a different formation timescale of the galactic halo.

According to calculations available in the literature, for [Fe/H] $> -1.4$ the majority of contribution on  the heaviest Mg isotopes comes from AGB stars with masses of about $4\pm1$M$_{\odot}$, which have a lifetime between of about 150-300 million years, which indicates a very short formation timescale of the galactic halo. We present a new halo model that reproduces the rapid increase of \iso{26}Mg/Mg ratios, by including a strong outflow. From the parameter study of the chemical evolution models, we conclude that the upper limit for the formation timescale is 1.5 Gyr.


\acknowledgments

M.C. would like to acknowledge support from CAPES. This work was also conducted during a scholarship supported by Capes/PDSE (88881.135113/2016-01) at Monash University. J.M. is thankful for the support of FAPESP (2012/24392-2, 2014/18100-4) and CNPq (Bolsa de Produtividade). The authors wish to recognize and acknowledge the very significant cultural role and reverence that the summit of Maunakea has always had within the indigenous Hawaiian community.  We are most fortunate to have the opportunity to conduct observations from this mountain.

%




\software{numpy \citep{numpy}, matplotlib \citep{hunter/07}, ATLAS9 \citep{castelli/04}, MOOG \citep{sneden73}, MAKEE (\url{http://www.astro.caltech.edu/~tb/makee/}), IRAF \citep{tody86,tody93}}.





\bibliographystyle{aasjournal}
\bibliography{mybib} 

\begin{thebibliography}{}
\expandafter\ifx\csname natexlab\endcsname\relax\def\natexlab#1{#1}\fi
\providecommand{\url}[1]{\href{#1}{#1}}

\bibitem[{{Alib{\'e}s} {et~al.}(2001){Alib{\'e}s}, {Labay}, \&
  {Canal}}]{alibes2001}
{Alib{\'e}s}, A., {Labay}, J., \& {Canal}, R. 2001, Astronomy \& Astrophysics,
  370, 1103

\bibitem[{{Arnett} \& {Thielemann}(1985)}]{arnett/85}
{Arnett}, W.~D., \& {Thielemann}, F.-K. 1985, \apj, 295, 589

\bibitem[{{Arnould} {et~al.}(1999){Arnould}, {Goriely}, \&
  {Jorissen}}]{arnould/99}
{Arnould}, M., {Goriely}, S., \& {Jorissen}, A. 1999, \aap, 347, 572

\bibitem[{{Asplund} {et~al.}(2009){Asplund}, {Grevesse}, {Sauval}, \&
  {Scott}}]{asplund09}
{Asplund}, M., {Grevesse}, N., {Sauval}, A.~J., \& {Scott}, P. 2009, \araa, 47,
  481

\bibitem[{{Barbuy}(1985)}]{barbuy85}
{Barbuy}, B. 1985, \aap, 151, 189

\bibitem[{{Barbuy}(1987)}]{barbuy87a}
---. 1987, \aap, 172, 251

\bibitem[{{Barbuy} {et~al.}(1987){Barbuy}, {Spite}, \& {Spite}}]{barbuy87b}
{Barbuy}, B., {Spite}, F., \& {Spite}, M. 1987, \aap, 178, 199

\bibitem[{{Carollo} {et~al.}(2007){Carollo}, {Beers}, {Lee}, {Chiba}, {Norris},
  {Wilhelm}, {Sivarani}, {Marsteller}, {Munn}, {Bailer-Jones}, {Fiorentin}, \&
  {York}}]{carollo07}
{Carollo}, D., {Beers}, T.~C., {Lee}, Y.~S., {et~al.} 2007, \nat, 450, 1020

\bibitem[{{Casagrande} {et~al.}(2010){Casagrande}, {Ram{\'{\i}}rez},
  {Mel{\'e}ndez}, {Bessell}, \& {Asplund}}]{casagrande10}
{Casagrande}, L., {Ram{\'{\i}}rez}, I., {Mel{\'e}ndez}, J., {Bessell}, M., \&
  {Asplund}, M. 2010, \aap, 512, A54

\bibitem[{{Castelli} \& {Kurucz}(2004)}]{castelli/04}
{Castelli}, F., \& {Kurucz}, R.~L. 2004, ArXiv Astrophysics e-prints,
  astro-ph/0405087

\bibitem[{{Chen} \& {Zhao}(2006)}]{chen06}
{Chen}, Y.~Q., \& {Zhao}, G. 2006, \mnras, 370, 2091

\bibitem[{{Chiappini} {et~al.}(1997){Chiappini}, {Matteucci}, \&
  {Gratton}}]{chiappini97}
{Chiappini}, C., {Matteucci}, F., \& {Gratton}, R. 1997, \apj, 477, 765

\bibitem[{{Chiba} \& {Yoshii}(1998)}]{chiba98}
{Chiba}, M., \& {Yoshii}, Y. 1998, \aj, 115, 168

\bibitem[{{Eggen} {et~al.}(1962){Eggen}, {Lynden-Bell}, \& {Sandage}}]{egen62}
{Eggen}, O.~J., {Lynden-Bell}, D., \& {Sandage}, A.~R. 1962, \apj, 136, 748

\bibitem[{{Fenner} {et~al.}(2003){Fenner}, {Gibson}, {Lee}, {Karakas},
  {Lattanzio}, {Chieffi}, {Limongi}, \& {Yong}}]{fenner03}
{Fenner}, Y., {Gibson}, B.~K., {Lee}, H.-c., {et~al.} 2003, PASA, 20, 340

\bibitem[{{Fishlock} {et~al.}(2014){Fishlock}, {Karakas}, {Lugaro}, \&
  {Yong}}]{fishlock14}
{Fishlock}, C.~K., {Karakas}, A.~I., {Lugaro}, M., \& {Yong}, D. 2014, \apj,
  797, 44

\bibitem[{{Gay} \& {Lambert}(2000)}]{gay00}
{Gay}, P.~L., \& {Lambert}, D.~L. 2000, \apj, 533, 260

\bibitem[{{Heger} \& {Woosley}(2010)}]{heger/10}
{Heger}, A., \& {Woosley}, S.~E. 2010, \apj, 724, 341

\bibitem[{Hunter(2007)}]{hunter/07}
Hunter, J.~D. 2007, Computing In Science \& Engineering, 9, 90

\bibitem[{{Karakas} \& {Lattanzio}(2003)}]{karakas03}
{Karakas}, A.~I., \& {Lattanzio}, J.~C. 2003, \pasa, 20, 279

\bibitem[{{Karakas} \& {Lattanzio}(2014)}]{karakas14}
---. 2014, \pasa, 31, e030

\bibitem[{{Kobayashi} {et~al.}(2011){Kobayashi}, {Karakas}, \&
  {Umeda}}]{kobayashi11}
{Kobayashi}, C., {Karakas}, A.~I., \& {Umeda}, H. 2011, MNRAS, 414, 3231

\bibitem[{{Kobayashi} \& {Nakasato}(2011)}]{kobayashi/nakasato/11}
{Kobayashi}, C., \& {Nakasato}, N. 2011, \apj, 729, 16

\bibitem[{{Kobayashi} {et~al.}(2000){Kobayashi}, {Tsujimoto}, \&
  {Nomoto}}]{kobayashi00}
{Kobayashi}, C., {Tsujimoto}, T., \& {Nomoto}, K. 2000, \apj, 539, 26

\bibitem[{{Kobayashi} {et~al.}(2006){Kobayashi}, {Umeda}, {Nomoto}, {Tominaga},
  \& {Ohkubo}}]{kobayashi06}
{Kobayashi}, C., {Umeda}, H., {Nomoto}, K., {Tominaga}, N., \& {Ohkubo}, T.
  2006, \apj, 653, 1145

\bibitem[{{Kroupa} {et~al.}(1993){Kroupa}, {Tout}, \& {Gilmore}}]{kroupa93}
{Kroupa}, P., {Tout}, C.~A., \& {Gilmore}, G. 1993, \mnras, 262, 545

\bibitem[{{McWilliam} \& {Lambert}(1988)}]{mcwilliam88}
{McWilliam}, A., \& {Lambert}, D.~L. 1988, \mnras, 230, 573

\bibitem[{{Mel{\'e}ndez} \& {Cohen}(2007)}]{melendez07}
{Mel{\'e}ndez}, J., \& {Cohen}, J.~G. 2007, \apjl, 659, L25

\bibitem[{{Mel{\'e}ndez} \& {Cohen}(2009)}]{melendez09}
---. 2009, \apj, 699, 2017

\bibitem[{{Micali} {et~al.}(2013){Micali}, {Matteucci}, \& {Romano}}]{micali13}
{Micali}, A., {Matteucci}, F., \& {Romano}, D. 2013, \mnras, 436, 1648

\bibitem[{{Monachesi} {et~al.}(2016){Monachesi}, {G{\'o}mez}, {Grand},
  {Kauffmann}, {Marinacci}, {Pakmor}, {Springel}, \& {Frenk}}]{monachesi/17}
{Monachesi}, A., {G{\'o}mez}, F.~A., {Grand}, R.~J.~J., {et~al.} 2016, \mnras,
  459, L46

\bibitem[{{Navarro} {et~al.}(1997){Navarro}, {Frenk}, \& {White}}]{navarro97}
{Navarro}, J.~F., {Frenk}, C.~S., \& {White}, S.~D.~M. 1997, \apj, 490, 493

\bibitem[{{Nissen} \& {Schuster}(2010)}]{nissen10}
{Nissen}, P.~E., \& {Schuster}, W.~J. 2010, \aap, 511, L10

\bibitem[{{Ram{\'{\i}}rez} \& {Mel{\'e}ndez}(2005)}]{ramirez05}
{Ram{\'{\i}}rez}, I., \& {Mel{\'e}ndez}, J. 2005, \apj, 626, 446

\bibitem[{{Searle} \& {Zinn}(1978)}]{searly78}
{Searle}, L., \& {Zinn}, R. 1978, \apj, 225, 357

\bibitem[{{Shingles} {et~al.}(2015){Shingles}, {Doherty}, {Karakas},
  {Stancliffe}, {Lattanzio}, \& {Lugaro}}]{shingles15}
{Shingles}, L.~J., {Doherty}, C.~L., {Karakas}, A.~I., {et~al.} 2015, \mnras,
  452, 2804

\bibitem[{{Sneden}(1973)}]{sneden73}
{Sneden}, C.~A. 1973, PhD thesis, THE UNIVERSITY OF TEXAS AT AUSTIN.

\bibitem[{{Soubiran} {et~al.}(2016){Soubiran}, {Le Campion}, {Brouillet}, \&
  {Chemin}}]{pastel16}
{Soubiran}, C., {Le Campion}, J.-F., {Brouillet}, N., \& {Chemin}, L. 2016,
  \aap, 591, A118

\bibitem[{{Spinrad} \& {Wood}(1965)}]{spinrad65}
{Spinrad}, H., \& {Wood}, D.~B. 1965, \apj, 141, 109

\bibitem[{{Thielemann} \& {Arnett}(1985)}]{thielemann/85}
{Thielemann}, F.~K., \& {Arnett}, W.~D. 1985, \apj, 295, 604

\bibitem[{{Thygesen} {et~al.}(2017){Thygesen}, {Kirby}, {Gallagher}, {Ludwig},
  {Caffau}, {Bonifacio}, \& {Sbordone}}]{thygesen17}
{Thygesen}, A.~O., {Kirby}, E.~N., {Gallagher}, A.~J., {et~al.} 2017, \apj,
  843, 144

\bibitem[{{Thygesen} {et~al.}(2016){Thygesen}, {Sbordone}, {Ludwig}, {Ventura},
  {Yong}, {Collet}, {Christlieb}, {Melendez}, \& {Zaggia}}]{thygesen16}
{Thygesen}, A.~O., {Sbordone}, L., {Ludwig}, H.-G., {et~al.} 2016, \aap, 588,
  A66

\bibitem[{{Tissera} {et~al.}(2012){Tissera}, {White}, \&
  {Scannapieco}}]{tissera/12}
{Tissera}, P.~B., {White}, S.~D.~M., \& {Scannapieco}, C. 2012, \mnras, 420,
  255

\bibitem[{{Tody}(1986)}]{tody86}
{Tody}, D. 1986, in \procspie, Vol. 627, Instrumentation in astronomy VI, ed.
  D.~L. {Crawford}, 733

\bibitem[{{Tody}(1993)}]{tody93}
{Tody}, D. 1993, in Astronomical Society of the Pacific Conference Series,
  Vol.~52, Astronomical Data Analysis Software and Systems II, ed. R.~J.
  {Hanisch}, R.~J.~V. {Brissenden}, \& J.~{Barnes}, 173

\bibitem[{{van der Walt} {et~al.}(2011){van der Walt}, {Colbert}, \&
  {Varoquaux}}]{numpy}
{van der Walt}, S., {Colbert}, S.~C., \& {Varoquaux}, G. 2011, Computing in
  Science \& Engineering, 13, 22

\bibitem[{{Ventura} \& {D'Antona}(2011)}]{ventura11}
{Ventura}, P., \& {D'Antona}, F. 2011, \mnras, 410, 2760

\bibitem[{{Vogt} {et~al.}(1994){Vogt}, {Allen}, {Bigelow}, {Bresee}, {Brown},
  {Cantrall}, {Conrad}, {Couture}, {Delaney}, {Epps}, {Hilyard}, {Hilyard},
  {Horn}, {Jern}, {Kanto}, {Keane}, {Kibrick}, {Lewis}, {Osborne},
  {Pardeilhan}, {Pfister}, {Ricketts}, {Robinson}, {Stover}, {Tucker}, {Ward},
  \& {Wei}}]{hires94}
{Vogt}, S.~S., {Allen}, S.~L., {Bigelow}, B.~C., {et~al.} 1994, in SPIE
  Proceedings, Vol. 2198, Instrumentation in Astronomy VIII, ed. D.~L.
  {Crawford} \& E.~R. {Craine}, 362

\bibitem[{{Woosley} \& {Weaver}(1995)}]{ww95}
{Woosley}, S.~E., \& {Weaver}, T.~A. 1995, \apjs, 101, 181

\bibitem[{{Yong} {et~al.}(2003{\natexlab{a}}){Yong}, {Grundahl}, {Lambert},
  {Nissen}, \& {Shetrone}}]{yong/grundahl03}
{Yong}, D., {Grundahl}, F., {Lambert}, D.~L., {Nissen}, P.~E., \& {Shetrone},
  M.~D. 2003{\natexlab{a}}, \aap, 402, 985

\bibitem[{{Yong} \& {Lambert}(2003)}]{yong03pasp}
{Yong}, D., \& {Lambert}, D.~L. 2003, PASP, 115, 22

\bibitem[{{Yong} {et~al.}(2004){Yong}, {Lambert}, {Allende Prieto}, \&
  {Paulson}}]{yong/lambert04}
{Yong}, D., {Lambert}, D.~L., {Allende Prieto}, C., \& {Paulson}, D.~B. 2004,
  \apj, 603, 697

\bibitem[{{Yong} {et~al.}(2003{\natexlab{b}}){Yong}, {Lambert}, \&
  {Ivans}}]{yong2003}
{Yong}, D., {Lambert}, D.~L., \& {Ivans}, I.~I. 2003{\natexlab{b}}, ApJ, 599,
  1357

\bibitem[{{Zolotov} {et~al.}(2009){Zolotov}, {Willman}, {Brooks}, {Governato},
  {Brook}, {Hogg}, {Quinn}, \& {Stinson}}]{zolotov/09}
{Zolotov}, A., {Willman}, B., {Brooks}, A.~M., {et~al.} 2009, \apj, 702, 1058

\end{thebibliography}



\end{document}